\documentclass{raa}            

\usepackage{graphicx,times}             

\begin{document}

 \title{An accreting low magnetic field magnetar for the ultraluminous X-ray source in M82
}

   \volnopage{Vol.0 (200x) No.0, 000--000}      
   \setcounter{page}{1}          

     \author{H. Tong
      \inst{}
      }

   \institute{Xinjiang Astronomical Observatory, Chinese Academy of Sciences, Urumqi, Xinjiang 830011,
    China; {\it tonghao@xao.ac.cn}\\
      }

   \date{Received~~2012 month day; accepted~~2012~~month day}

\abstract{ 
One ultraluminous X-ray source in M82 is identified as an accreting neutron star recently
(named as NuSTAR J095551+6940.8).
It has a super-Eddington luminosity and is spinning up. For an aged magnetar, 
it is more likely to be a low magnetic field magnetar. An accreting low magnetic field magnetar
may explain both the super-Eddington luminosity and the rotational behavior of this source.  
Considering the effect of beaming, the spin-up rate is understandable using the traditional form of accretion torque. 
The transient nature, spectral properties of M82 X-2 are discussed. 
The theoretical period range of accreting magnetar is provided. 
Three observational appearances of accreting magnetars are summarized. 
}

   \authorrunning{H. Tong}            
   \titlerunning{Accreting low-B magnetar for the ULX in M82}  

   \maketitle

%
%
\section{Introduction}

Pulsars are rotating magnetized neutron stars. Up to now, various kinds of pulsar-like 
objects are discovered (Tong \& Wang 2014). Among them are: normal pulsars whose surface dipole field
is about $10^{12} \,\rm G$ (e.g., the Crab pulsar, Wang et al. 2012); high magnetic field pulsars 
with surface dipole field as high as $10^{14}\,\rm G$ (Ng \& Kaspi 2010); central compact objects's surface 
magnetic field is at the lower end, about $10^{10} \,\rm G$ (Gotthelf et al. 2013). Millisecond pulsars
are thought to recycled neutron stars (Alpar et al. 1982). Their surface dipole fields may have decreased significantly 
during the recycling process (Zhang \& Kojima 2006), which can be as low as a few times $10^8 \,\rm G$.
Magnetars are thought to be neutron stars powered by their strong magnetic fields (Duncan \& Thompson 1992). 
Their surface dipole fields can be as high as $10^{14} -10^{15}\,\rm G$ (Tong et al. 2013). At the same time, 
they may have even higher multipole fields (Tong \& Xu 2011, 2014). For an aged magnetar, its dipole magnetic 
field may have decreased a lot ($\sim 10^{12} \,\rm G$, Turolla et al. 2011). At the same time,  
their surface multipole fields may still in the magnetar range (i.e., ``low magnetic field'' magnetar). 
Several low magnetic field magnetars are known (Rea et al. 2010, 2012; Zhou et al. 2014).

Accretion powered X-ray pulsars are discovered at the beginning of X-ray astronomy. 
Since magnetars are just a special kind of neutron stars, an accreting magnetar is also expected. 
However, no strong observational evidence for the existence of an accreting magnetar is found (Wang 2013; Tong \& Wang 2014). 
Possible observational signatures of accreting magnetars are discussed in Tong \& Wang (2014), 
including magnetar-like bursts, and a hard X-ray tail etc. The recently discovered ultraluminous 
X-ray pulsar in M82 (NuSTAR J095551+6940.8, Bachetti et al. 2014) may be another manifestation of accreting magnetars. 

Ultraluminous X-ray sources are commonly assumed to accreting black holes (with either stellar mass or intermediate 
mass, Liu et al. 2013; Feng \& Soria 2011). 
The discovery of pulsation period and spin-up trend of an ultraluminous X-ray source 
in M82 points to an accreting neutron star (Bachetti et al. 2014). The neutron star's X-ray luminosity can be 
as high as $10^{40} \,\rm erg \,s^{-1}$, with rotational period $1.37\,\rm s$ and 
period derivative $\dot{P} \approx -2\times 10^{-10}$ (Bachetti et al. 2014). If the central neutron star is a 
low magnetic field magnetar, an accreting low magnetic field magnetar may explain 
both the radiative and timing observations.

Model calculations are presented in Section 2, including super-Eddington luminosity (Section 2.1) 
and rotational behaviors (Section 2.2). 
Discussion and conclusion are given in Section 3 and Section 4, respectively.

\section{Accreting low magnetic field magnetar}

\subsection{Super-Eddington luminosity}

According to Bachetti et al. (2014), the ultraluminous X-ray pulsar NuSTAR J095551+6940.8 has a pulsed luminosity
$4.9\times 10^{39} \,\rm erg \,s^{-1}$ (in the energy range $3$-$30\,\rm keV$). 
While there are more than one ultraluminous X-ray sources in M82 (Kaaret et al. 2006). According to the 
centroid of pulsed flux, the ultraluminous X-ray source M82 X-2 may be the counterpart of NuSTAR J095551+6940.8. 
Soft X-ray observation of M82 X-2 shows the luminosity is $6.6\times 10^{39} \,\rm erg \,s^{-1}$ 
(in the energy range $0.5$-$10 \,\rm keV$). Therefore, the total X-ray luminosity of NuSTAR J095551+6940.8 may be 
(assuming isotropic emission, Bachetti et al. 2014)
\begin{equation}
 L_{\rm iso} (0.5-30\,{\rm keV})= L_{\rm iso,40} \times 10^{40} \,\rm erg \,s^{-1},
\end{equation}
where $L_{\rm iso,40} \approx 1$. For an accreting neutron star, the dipole magnetic field will channel the accretion 
material into columns near the star's polar cap (Sahpiro \& Teukolsky 1983). Therefore, the emission of the neutron star
is expected to be beamed (Gnedin \& Sunyaev 1973). The true X-ray luminosity should be corrected by a beaming factor 
\begin{equation}\label{X-ray luminosity}
 L_{\rm x} (0.5-30 \,{\rm keV}) = b\, L_{\rm iso} =b \, L_{\rm iso, 40} \times 10^{40} \,\rm erg \,s^{-1},
\end{equation}
where $b<1$ is beaming factor. From previous pulse profile observations of accreting neutron stars 
(figure 7 in Bildsten et al. 1997), there should be some amount of beaming\footnote{The ultraluminous X-ray pulsar 
NuSTAR J095551+6940.8 also has some pulse profile information, see figure 1 in Bachetti et al. 2014.}. 
If the ducy cycle of the pulse profile is about $50\%$, then the solid angle of the radiation beam may only occupy $25\%$ 
of the whole sky\footnote{This is a very crude estimation. In principle, the beaming factor adopted in the following 
is an assumption.}. 
In the following, a beaming factor of $b=0.2$ is chosen 
(or $b^{-1} =5$, consistent with other observational constraints, Feng \& Soria 2011).

\subsubsection{Accreting normal neutron star}

The maximum luminosity for steady spherical accretion is (i.e., the Eddington limit, Frank et al. 2002)
\begin{equation}
 L_{\rm Edd} = 1.3\times 10^{38} M_1 \,\rm erg \,s^{-1},
\end{equation}
where $M_1$ is the mass of the central star in units of solar masses. 
Considering the modification due to accretion column,
the maximum luminosity for an accreting neutron star is several times higher 
(Basko \& Sunyaev 1976, denoted as critical luminosity in the following)
\begin{equation}
 L_{\rm cr} = \frac{l_0}{2\pi d_0} L_{\rm Edd} = 8\times 10^{38} \left( \frac{l_0/d_0}{40} \right) M_1 \,\rm erg \,s^{-1},
\end{equation}
where $l_0$ is the length of the accretion column, $d_0$ is the thickness. 
The typical value of $l_0/d_0$ is about $40$ (Basko \& Sunyaev 1976). Both theory and observation of neutron stars show that
they may have a mass in excess of $1.4$ solar masses (e.g., $2$ solar masses) (Lai \& Xu 2011 and references therein). 
The existence of two solar mass neutron stars may be difficult to understand compared with other neutron star mass measurement
(Zhang et al. 2011). One way to form heavy neutron stars may involve super-Eddington accretion (Lee \& Cho 2014). 
Since NuSTAR J095551+6940.8 is probably accreting at 
super-Eddington rate, it may also have a larger mass.
If the central neutron star is a massive one\footnote{The following conlusions are unaffected by a different choice of central 
neutron star mass, e.g., $M_1 =1.4$.} with $M_1 =2$, 
the theoretical maximum luminosity is $L_{\rm cr} =1.6\times 10^{39} (\frac{l_0/d_0}{40})\,\rm erg \,s^{-1}$. 
For a beaming factor  $b=0.2$, the true X-ray luminosity is $L_{\rm x} =2\times 10^{39} L_{\rm iso,40} \,\rm erg \,s^{-1}$. 
Therefore, it cannot be ruled out that the central neutron star of NuSTAR J095551+6940.8 is a massive neutron star (with no peculiarity in the 
magnetic properties). Meanwhile, for an accreting massive neutron star, the maximum apparent isotropic luminosity 
will be in the range $10^{40} \,\rm erg \,s^{-1}$. It is very hard to reach a luminosity higher than $10^{40} \,\rm erg \,s^{-1}$. 
In this case, NuSTAR J095551+6940.8 will be the extreme example of accreting normal neutron stars. 

\subsubsection{Accreting magnetar}

The super-Eddington luminosity is easier to understand in the magnetar case. Magnetars can have giant flares due to sudden release 
of magnetic energy. In the pulsating tail, the star's luminosity can be as high as $10^{42} \,\rm erg \,s^{-1}$, lasting for
about hundreds of seconds (Mereghetti 2008). One of the reasons to propose the magnetar idea was to explain this super-Eddington 
luminosity (Paczynski 1992). The same argument can also be applied to the ultraluminous X-ray pulsar in M82.
The scattering cross section between electrons and photons is significantly suppressed in the presence of strong magnetic field 
(only for one polarization). In order to obtain the corresponding critical luminosity, some average (e.g., Rosseland mean) of
cross section (or opacity) is needed. The final result is (Paczynski 1992)
\begin{equation}
 \frac{L_{\rm cr}}{L_{\rm Edd}} \approx 2 \times \left( \frac{B}{10^{12} \,\rm G} \right)^{4/3},
\end{equation}
which is only valid for $L_{\rm cr} \gg L_{\rm Edd}$. 
If the total magnetic field near the polar cap\footnote{Here only the total magnetic field strength 
near the polar cap is required. No specific magnetic field geometry is assumed.} is $10^{14} \,\rm G$, 
then the critical luminosity is $L_{\rm cr} \approx 10^3 L_{\rm Edd} \approx 10^{41} \,\rm erg \,s^{-1}$.
Considering the geometry 
of accretion column, the critical luminosity may be even higher (Basko \& Sunyaev 1976). 
In the accreting magnetar case, even the most luminous sources with a luminosity as high as $10^{41} \,\rm erg \,s^{-1}$
are possible. Therefore, the ultraluminous X-ray pulsar in M82 with 
isotropic luminosity about $10^{40} \,\rm erg \,s^{-1}$ can be safely understood in the accreting magnetar case.

\subsection{Rotational behaviors}

The ultraluminous X-ray pulsar NuSTAR J095551+6940.8 has a rotational period $P=1.37 \,\rm s$ (Bachetti et al. 2014). 
At the same time, the pulsar is spinning up (i.e., the rotational period is decreasing). 
The period derivative is roughly about $\dot{P} \approx -2\times 10^{-10}$ (Bachetti et al. 2014). 
For this accreting neutron star, its light cylinder radius is (where the rotational velocity equals the 
speed of light) $R_{\rm lc} = \frac{P c}{2\pi} = 6.5\times 10^{9} \,\rm cm$. The corotation radius is
defined as where the local Keplerian velocity equals the rotational velocity
\begin{equation}
 R_{\rm co} = \left( \frac{G M}{4\pi^2} \right)^{1/3} P^{2/3} = 1.8\times 10^{8} M_1^{1/3} \,\rm cm,
\end{equation}
where $G$ is the gravitational constant. In the presence of magnetic field, the accretion flow will be
controlled by the magnetic field. The Alfv$\acute{e}$n radius characterize this quantitatively. It is defined as the 
radius where the magnetic energy density equals the kinetic energy density of the accretion flow 
(Shapiro \& Teukolsky 1983; Lai 2014)
\begin{equation}\label{Alfven radius}
 R_{\rm A} = 3.2\times 10^8 \, \mu_{30}^{4/7} M_1^{-1/7} \dot{M}_{17}^{-2/7} \,\rm cm,
\end{equation}
where $\mu_{30}$ is the dipole magnetic moment in units of $10^{30} \,\rm G \,cm^3$, $\dot{M}_{17}$ is the 
mass accretion rate in units of $10^{17} \,\rm g \,s^{-1}$ (the corresponding luminosity is about $10^{37} \,\rm erg \,s^{-1}$).
When the Alfv$\acute{e}$n radius is smaller than the light cylinder radius, the accretion flow may interact 
with the central neutron star. In the case of spin equilibrium, the Alfv$\acute{e}$n radius is equal to the 
corotation radius (Lai 2014). NuSTAR J095551+6940.8 may be in spin equilibrium ($| P/\dot{P}| \approx 200$ years). 
However, its counterpart M82 X-2 is a transient source (Feng et al. 2007; Kong et al. 2007). 
Therefore, whether it is in spin equilibrium is not certain (i.e., which luminosity corresponds to the spin equilibrium case 
is not known). The period derivative measurement of this source 
means that the star is experiencing some accretion torque. From this point, the star's dipole magnetic field 
may be determined. Whether the star is in spin equilibrium can be checked consequently. 

For the X-ray luminosity in equation (\ref{X-ray luminosity}), the corresponding accretion rate onto the neutron star is 
\begin{equation}
 \dot{M}_{\rm acc} = \frac{R}{G M} L_{\rm x} =7.5\times 10^{19} b L_{\rm iso, 40} R_6 M_1^{-1} \,\rm g\,s^{-1},
\end{equation}
where $M$ is the mass of the neutron star, $R$ is the radius of the neutron star , and $R_6$ is the radius in units of 
$10^{6} \,\rm cm$. The corresponding  Alfv$\acute{e}$n radius is 
\begin{equation}\label{Alfven radius 2}
 R_{\rm A} = 4.8\times 10^7 \mu_{30}^{4/7} M_1^{1/7} (b L_{\rm iso, 40} R_6)^{-2/7} \,\rm cm.
\end{equation}
The angular momentum carried onto the neutron star by the accretion matter is (Shapiro \& Teukolsky 1983, which follows 
the treatment of Ghosh \& Lamb 1979; Lai 2014): $\dot{M}_{\rm acc} \sqrt{G M R_{\rm A}}$. The angular momentum of 
central neutron star is $J = I \Omega $, where $I=2/5 M R^2$ is the moment of inertia of the neutron star, 
$\Omega=2\pi/P$ is the angular velocity. The change of stellar angular momentum is 
$\dot{J} = I \dot{\Omega} = -2\pi I \dot{P}/P^2$ (the change of moment of inertia is negliable, Shaprio \& Teukolsky 1983).
According to the conservation of angular momentum 
\begin{equation}\label{spin-up rate}
 -2\pi I \frac{\dot{P}}{P^2} = \dot{M}_{\rm acc} \sqrt{G M R_{\rm A}}.
\end{equation}
Therefore, the dipole magnetic moment of the neutron star in NuSTAR J095551+6940.8 is 
\begin{equation}\label{magnetic moment}
 \mu_{30} = 2\times 10^{-4} M_1^5 R_6^4 b^{-3} L_{\rm iso, 40}^{-3}.
\end{equation}
The dipole magnetic moment is related with the polar magnetic field as $\mu =1/2 B_{\rm p} R^3$ (Shapiro \& Teukolsky 1983; Tong et al. 2013). 
The corresponding magnetic field at the neutron star polar cap is 
\begin{equation}
 B_{\rm p} =4 \times 10^8 M_1^5 R_6 b^{-3} L_{\rm iso, 40}^{-3} \,\rm G.
\end{equation}
For a two solar mass neutron star ($M_1=2$) with a beaming factor $b=0.2$, the dipole magnetic field 
is about $B_{\rm p} =1.6\times 10^{12} R_6 L_{\rm iso, 40}^{-3} \,\rm G$. Combined with super-Eddington luminosity
requirement, the central neutron star is likely to be a low magnetic field magnetar. The star's high multipole field
near the surface (about $10^{14} \,\rm G$) accounts for the super-Eddington luminosity. The much lower dipole field 
(about $10^{12} \,\rm G$) is responsible for the rotational behaviors. From an evolution point of view, an aged magnetar
is also more like to be a low magnetic field magnetar (Turolla et al. 2011). 

Since M82 X-2 (the possbile counterpart of NuSTAR J095551+6940.8) is highly variable. Its peak luminosity can reach 
$2.2\times 10^{40} \,\rm erg s^{-1}$ (Feng et al. 2007). In its low state, the source is below the detection limit, 
with luminosity lower than $10^{37}$-$10^{38}\,\rm erg \,s^{-1}$ (Feng et al. 2007; Kong et al. 2007, different 
authors gave different estimations). Whether NuSTAR J095551+6940.8 is in spin equilibrium is detemined by the long 
term average mass accretion rate, which unfortunately is not known precisely at present. Considering the variation 
of X-ray luminosity, the average accretion rate can be in the range $10^{17}$-$10^{20} \,\rm g \,s^{-1}$. 
The equilibrium period can be determined by equaling the corotation radius and the Alfv$\acute{e}$n radius (Lai 2014)
\begin{equation}\label{equalibrium period}
 P_{\rm eq} = 3.1 \mu_{30}^{6/7} M_1^{-5/7} \dot{M}_{\rm ave,17}^{-3/7} \,\rm s,
\end{equation}
where $\dot{M}_{\rm ave, 17}$ is average accretion rate in units of $10^{17} \,\rm g \,s^{-1}$. 
Substituting the magnetic moment in equation (\ref{magnetic moment}), the equilibrium period of NuSTAR J095551+6940.8 is 
$P_{\rm eq} = 2\times 10^{-3} M_1^{25/7} R_6^{24/7} b^{-18/7} L_{\rm iso, 40}^{-18/7} \dot{M}_{\rm ave,17}^{-3/7} \,\rm s$. 
For typical parameters, $M_1=2$ and $b=0.2$, the corresponding equilibrium period is 
$P_{\rm eq} = 1.6 R_6^{24/7} L_{\rm iso, 40}^{-18/7} \dot{M}_{\rm ave,17}^{-3/7} \,\rm s$. 
If the long term average accretion rate of NuSTAR J095551+6940.8 is approximately $10^{17} \,\rm g \,s^{-1}$, 
then it may be in spin equilibrium (with current period of $1.37\,\rm s$). If the long term average accretiong rate 
is $10^2$ ($10^3$) times higher, the equilibrium period will be about $0.2\,\rm s$ ($0.1\,\rm s$). 
Then the neutron star is not in spin equilibrium and should experience some kind of net spin up. 
This is also consistent with observations (with period derivative $-2\times 10^{-10}$). 
In the current status, both of these two cases are possible. 

\section{Discussion}

\subsection{Transient nature}

If M82 X-2 is indeed the counterpart of NuSTAR J095551+6940.8, then more information is available.
M82 X-2 is a transient source. Its luminosity ranges from $10^{40} \,\rm erg \,s^{-1}$ to lower than 
$10^{37}$-$10^{38}\,\rm erg \,s^{-1}$ (Feng et al. 2007; Kong et al. 2007). One possibility
is that the neutron star switches between accretion phase and propeller phase (Cui 1997). If the neutron star is 
near spin equilibrium (the Alfv$\acute{e}$n radius is approximately equal to the corotation radius), 
a higher accretion rate will result in a higher X-ray luminosity (accretion phase and spin-up). When the accretion 
rate is lower, the Alfv$\acute{e}$n radius will be larger (see equation (\ref{Alfven radius})). 
Then the centrifugal force will be larger than gravitational force. 
The accretion matter that can fall onto the neutron star will be greatly reduced (propeller phase and spin-down). 
A much lower X-ray luminosity is expected in the propeller phase, as have
been observed in other accreting neutron star systems (Cui 1997; Zhang et al. 1998). 
The transient nature of M82 X-2 may due to switches between the accretion phase and the propeller phase. 

\subsection{Spectra properties}

There may be a disk component in the soft X-ray spectra of M82 X-2 (at $4.1\sigma$ significance level, Feng et al. 2010). 
The inner disk radius is about $3.5_{-1.9}^{+3.0} \times 10^9 \,\rm cm$ ($90\%$ confidence level). 
The inner disk temperature is about $0.17\pm 0.03 \,\rm keV$ (Feng et al. 2010). 
According to the above calculations, the typical Alfv$\acute{e}$n radius is about 
$7.5\times 10^7 R_6^2 L_{\rm iso, 40}^{-2} \,\rm cm$ (by substituting equation (\ref{magnetic moment})
into equation (\ref{Alfven radius 2})). For the standard thin disk, the disk temperature 
at the Alfv$\acute{e}$n radius is about $0.15 \,\rm keV$ (using equation (5.43) in Frank et al. 2002). 
For an accreting neutron star the Alfv$\acute{e}$n radius may be the inner disk radius (Lai 2014). 
However, the observed inner disk radius is very uncertain. The theoretical temperature at the Alfv$\acute{e}$n 
radius is consistent with the observed inner disk temperature. 
Future more accurate determination of the disk radius may constrain this model 
(and other models, see below).

\subsection{Period range of accreting magnetars}

From equation (\ref{equalibrium period}), the equilibrium period ranges from about $0.1 \,\rm s$ to several seconds 
for an accreting low magnetic field mangetar (with dipole field about $10^{12} \,\rm G$). The exact value is determined 
by the long term average mass accretion rate. If for some accreting magnetars their surface dipole field is still 
very high (the extreme value is $10^{15} \,\rm G$ ), then the corresponding the equilibrium period can be as high as 
$10^3 \,\rm s$. Therefore, the period range of accreting magnetars may range from $0.1 \,\rm s$ to $10^{3} \,\rm s$. 
If the orbital period is about several days as in the case of NuSTAR J095551+6940.8, the X-ray observation 
time scale (tens of kiloseconds) will be a significant fraction of the orbital period. 
Accelerated searching technique must be employed in order to find out these periodic pulsations 
(Bachetti et al. 2014). 

\subsection{Observational appearances of accreting magnetars}

The discovery of low magnetic field magnetars (with a dipole field a few times $10^{12} \,\rm G$, 
Rea et al. 2010,2012; Zhou et al. 2014; Tong \& Xu 2012, 2013) clearly demonstrates that the multipole field is 
the crucial ingredient of magnetars. In order to power both the persistent emission and bursts, 
the dipole field is not enough. Stronger multipole field (about or higher than $10^{14} \,\rm G$) is needed.
Several failed predictions of the magnetar model (the supernova energy associated with magnetars is of normal value, 
the non-detection of magnetars by the Fermi telescope etc) have challenged the existence of strong dipole field 
in magnetars (Tong \& Xu 2011 and references therein). It is shown that magnetars may be wind braking and 
a strong dipole magnetic field is not necessary (Tong et al. 2013). The key ingredient of magnetars is their 
strong multipole field.
Signature of strong multipole field is needed in order to say that an accreting magnetar is observed
(Tong \& Wang 2014). From equation (\ref{equalibrium period}), 
for an accreting high magnetic field neutron star (with dipole field higher than $10^{14} \,\rm G$) the 
equilibrium period will be larger than one hundred seconds. Previously, some super-slow X-ray pulsars 
are thought to be accreting magnetars (with pulsation period longer than $10^3 \,\rm s$, Wang 2013). 
However, this is at most the observational evidence of strong dipole field. A neutron star with 
a strong dipole field is not a magnetar (Ng \& Kaspi 2010). 
Tong \& Wang (2014) had discussed possible observational appearances of accreting magnetars.
Combined with the result in this paper, three observational appearances of accreting magnetars 
are available at present: (1) magnetar-like bursts, (2) a hard X-ray tail (higher than $100 \,\rm keV$), and
(3) an ultraluminous  X-ray pulsar. 


\subsection{Comparison with other papers}

In the observational paper, Bachetti et al. (2014) made some estimations and showed that it 
may be difficult to explain both the super-Eddington luminosity and the spin-up rate. Assuming spin 
equilibrium, the Alfv$\acute{e}$n radius will be approximately equal to the corotation radius. Not considering 
the effect of beaming, the luminosity $10^{40} \,\rm erg \,s^{-1}$ requires a mass accretion rate 
about $10^{20} \,\rm g\,s^{-1}$. According to equation (\ref{spin-up rate}), the theoretical spin-up 
rate is about $-6\times 10^{-9}$. The observed spin-up rate is only about $-2\times 10^{-10}$. 
In order to solve this controversy, Eksi et al. (2014) and Lyutikov (2014) tried different forms of accretion torque. 
However, according to the above calculations, the observed spin-up rate 
is understandable even in the traditional formula of accretion torque provided that 
the effect of beaming is considered. With only one ultraluminous X-ray pulsar at hand, there are many
uncertainties. More observations of more sources are needed in order to make clear this problem. 

\section{Conclusion}

The ultraluminous X-ray pulsar NuSTAR J095551+6940.8 in M82 is modeled as an accreting low magnetic field 
magnetar. The magnetar strength multipole field is responsible for the super-Eddington 
luminosity. The much lower large scale dipole field determines the interaction between the neutron star
and the accretion flow. Its rotational behaviors can be explained using traditional 
form of accretion torque considering the effect of beaming. The counterpart of NuSTAR J095551+6940.8 (M82 X-2) is a transient because it may switch
between accretion phase and propeller phase. The theoretical period range of accreting magnetars may be 
very wide. Three observational appearances of accreting magnetar are available at present.

\section*{Acknowledgments}
The author would like to thank W. Wang, R. X. Xu, X. D. Li, J. F. Liu,  C. H. Lee, K. J. Lee, and Z. B. Li for discussions.
H.Tong is supported by Xinjiang Bairen project, NSFC (11103021), West Light Foundation of CAS (LHXZ201201), 
Qing Cu Hui of CAS, and 973 Program (2015CB857100).


\begin{thebibliography}{99}

\bibitem{Alpar1982}
Alpar, M. A., Cheng, A. F., Ruderman, M. A., et al. 1982, Nature, 300, 728

\bibitem{Bachetti2014}
Bachetti, M., Harrison, F. A., Walton, W. J., et al. 2014, Nature, 514, 202	

\bibitem{Basko1976}
Basko, M. M., \& Sunyaev, R. A. 1976, MNRAS, 175, 395	

\bibitem{bil97} 
Bildsten, L., Chakrabarty, D., Chiu, J., et al. 1997, ApJS, 113, 367

\bibitem{Cui1997}
Cui, W. 1997, ApJ, 482, L163

\bibitem{DT1992}
Duncan, R. C., \& Thompson, C. 1992, ApJ, 392, L9

\bibitem{Eksi2014}
Eksi, K. Y., Andac, I. C., Cikintoglu, S., et al. 2014, arXiv:1410.5205

\bibitem{Feng2007}
Feng, H., \& Kaaret, P. 2007, ApJ, 668, 941 

\bibitem{Feng2010}
Feng, H., Rao, F., \& Kaaret, F. 2010, ApJ, 710, L137	

\bibitem{Feng2011}
Feng, H., \& Soria, R. 2011, New Astronomy Reviews, 55, 166

\bibitem{Frank2002}
Frank, J., King, A., Raine, D. 2002, Accretion power in astrophysics, Cambridge University Press, New York

\bibitem{Ghosh1979}
Ghosh, P., \& Lamb, F. K. 1979, ApJ, 234, 296

\bibitem{Gnedin1973}
Gnedin, Y. N., \& Sunyaev, R. A. 1973, A\&A, 25, 233

\bibitem{Gotthelf2013}
Gotthelf, E. V., Halpern, J. P., \& Alford, J. 2013, ApJ, 765, 58

\bibitem{Kaaret2006}
Kaaret, P., Simet, M. G., \& Lang, C. C. 2006, ApJ, 646, 174

\bibitem{Kong2007}
Kong, A. K. H., Yang, Y. J., Hsieh, P. Y., et al. 2007, ApJ, 671, 349 

\bibitem{Lai2014}
Lai, D. 2014, arXiv:1402.1903

\bibitem{Lai2011}
Lai, X. Y., \& Xu, R. X. 2011, RAA, 11, 687

\bibitem{Lee2014}
Lee, C. H., \& Cho, H. S. 2014, NuPhA, 928, 296

\bibitem{Liu2013}
Liu, J. F., Bergman, J. N., Bai, Y., et al. 2013, Nature, 503, 500

\bibitem{Lyutikov2014}
Lyutikov, M. 2014, arXiv:1410.8745

\bibitem{Mereghetti (2008)}
Mereghetti, S. 2008, A\&ARv, 15, 225

\bibitem{Ng2010}
Ng, C. Y., \& Kaspi, V. M. 2010, arXiv:1010.4592

\bibitem{Paczynski1992}
Paczynski, B. 1992, ACTA ASTRONOMICA, 42, 145

\bibitem{Rea2010}
Rea, N., Esposito, P., Turolla, R., et al. 2010, Science, 330, 944

\bibitem{Rea2012}
Rea, N., Israel, G. L., Esposito, P., et al. 2012, ApJ, 754, 27

\bibitem{Shapiro (1983)}
Shapiro, S. L., \& Teukolsky S. A. 1983, Black holes, white dwarfs, and neutron stars, John Wiley \& Sons, New York

\bibitem{Tong2011}
Tong, H., \& Xu, R. X. 2011, Int.Jour.Mod. Phys. E,20, 15 

\bibitem{Tong2012}
Tong, H., \& Xu, R. X. 2012, ApJ, 757, L10

\bibitem{Tong2013a}
Tong, H., \& Xu, R. X. 2013, RAA, 13, 1207

\bibitem{Tong2013}
Tong, H., Xu, R. X., Song, L. M., \& Qiao, G. J. 2013, ApJ, 768,144

\bibitem{TongWang2014}
Tong, H., \& Wang, W. 2014, arXiv:1406.6458

\bibitem{Tong2014}
Tong, H., \& Xu, R. X. 2014, AN, 335, 757

\bibitem{Turolla2011}
Turolla, R., Zane, S., Pons, J. A., et al. 2011, ApJ, 740, 105

\bibitem{Wang2012}
Wang, J., Wang, N., Tong, H., \& Yuan, J. 2012, Ap\&SS, 340, 307 

\bibitem{Wang2013}
Wang, W. 2013, Proceeding of IAUS, 291, 203 (arXiv:1211.5214)

\bibitem{Zhang2006}
Zhang, C. M., \& Kojima, Y. 2006, MNRAS, 366, 137

\bibitem{Zhang2011}
Zhang, C. M., Wang, J., Zhao, Y. H., et al. 2011, A\&A, 527, 83

\bibitem{Zhang1998}
Zhang, S. N., Yu, W., \& Zhang, W. 1998, ApJ, 494, L71

\bibitem{Zhou2014}
Zhou, P., Chen, Y., Li, X. D., et al. 2014, ApJ, 781, L16


\end{thebibliography}
\end{document}